\documentclass[sagev,shortAfour,times]{sagej}
\usepackage{multirow}
\usepackage{setspace}
\usepackage{bm}

\title{A Bayesian response-adaptive dose finding and comparative effectiveness trial}
\author{Heath, A.\affilnum{1, 2, 3}, 
Yaskina, M.\affilnum{4},
 Pechlivanoglou, P.\affilnum{1, 5}, 
 Rios, J.\affilnum{1},
 Offringa, M.\affilnum{1, 5},
Klassen, T.P \affilnum{6, 7},
 Poonai, N.\affilnum{8, 9} and
Pullenayegum, E.\affilnum{1, 2}} 

\affiliation{\affilnum{1}Child Health Evaluative Sciences, The Hospital for Sick Children \\
\affilnum{2} Division of Biostatistics, University of Toronto \\
\affilnum{3} Department of Statistical Science, University College London \\
\affilnum{4} Women \& Children's Health Research Institute, University of Alberta \\
\affilnum{5} Institute of Health Policy, Management and Evaluation, University of Toronto\\
\affilnum{6} University of Manitoba \\
\affilnum{7} Children’s Hospital Research Institute of Manitoba \\
\affilnum{8} Schulich School of Medicine and Dentistry \\
\affilnum{9} Children's Health Research Institute, London Health Sciences Centre } 

\runninghead{Heath, et al.}
\corrauth{Anna Heath, Peter Gilgan Centre for Research and Learning, 686 Bay Street, Toronto, ON, M5G 0A4}
\email{anna.heath@sickkids.ca}

\begin{document}

\maketitle

\vspace{-3.5cm}

\section*{Abstract}
\textbf{Background/Aims}: Combinations of treatments that have already received regulatory approval can offer additional benefit over each of the treatments individually. However, trials of these combinations are lower priority than those that develop novel therapies, which can restrict funding, timelines and patient availability. This paper develops a novel trial design to facilitate the evaluation of new combination therapies. This trial design combines elements of phase II and phase III trials to reduce the burden of evaluating combination therapies, whilst also maintaining a feasible sample size. This design was developed for a randomised trial that compares the properties of three combination doses of ketamine and dexmedetomidine, given intranasally, to ketamine delivered intravenously for children undergoing a closed reduction for a fracture or dislocation. 

\textbf{Methods}: This trial design uses response adaptive randomisation to evaluate different dose combinations and increase the information collected for successful novel drug combinations. The design then uses Bayesian dose-response modelling to undertake a comparative-effectiveness analysis for the most successful dose combination against a relevant comparator. We used simulation methods to determine the thresholds for adapting the trial and making conclusions. We also used simulations to evaluate the probability of selecting the dose combination with the highest true effectiveness, the operating characteristics of the design and its Bayesian predictive power. 

\textbf{Results}: With 410 participants, 5 interim updates of the randomisation ratio and a probability of effectiveness of 0.93, 0.88 and 0.83 for the three dose combinations, we have an 83\% chance of randomising the largest number of patients to the drug with the highest probability of effectiveness. Based on this adaptive randomisation procedure, the comparative effectiveness analysis has a type I error of less than 5\% and a 93\% chance of correctly concluding non-inferiority when the probability of effectiveness for the optimal combination therapy is 0.9. In this case, the trial has a 77\% chance of meeting its dual aims of dose finding and comparative effectiveness. Finally, the Bayesian predictive power of the trial is over 90\%. 

\textbf{Conclusions}: By simultaneously determining the optimal dose and collecting data on the relative effectiveness of an intervention, we can minimise administrative burden and recruitment time for a trial. This will minimise the time required to get effective, safe combination therapies to patients quickly. The proposed trial has high potential to meet the dual study objectives within a feasible overall sample size.

\paragraph{Keywords} Response adaptive trial, Non-inferiority trial, Bayesian Analysis, Clinical Trial Design

\section*{Introduction}
Investigator-initiated trials, where clinician investigators undertake their own trials,\cite{Suvarna:2012} are key to expanding the use of therapies that have received regulatory approval.\cite{Konwaretal:2018} One key expansion develops therapies that combine two or more currently available interventions to improve outcomes compared to either treatment alone.\cite{Mokhtarietal:2017, Kelleheretal:2018, Hartlingetal:2016} To develop these novel combination therapies, we must determine the optimal combination and evaluate its comparative effectiveness to the current standard of care. 

In standard drug development processes, these two aims require two trials; a phase II trial to determine the optimal dose combination and a phase III trial to evaluate the comparative effectiveness.\cite{KnoopWorden:1988} However, initiating a trial is time and resource intensive,\cite{Abramsetal:2013} especially when separate funding must be sought for each phase. Thus, trial designs that incorporate both of these elements, whilst maintaining a reasonable level of patient recruitment, can improve the efficiency of the trial process.\cite{StallardTodd:2011} 

Seamless phase II/III trials that move between the investigative Phase II and the confirmatory Phase III are the most common type of adaptive trials \cite{Bothwelletal:2018} as they are more efficient than two independent trials.\cite{Bretzetal:2006} These trials typically combine two distinct phases, where successful treatments are `taken forward' from Phase II to Phase III.\cite{StallardTodd:2011} Thus, statistical designs are available to select the optimal treatment from a set of alternatives before undertaking a comparative effectiveness analysis between the optimal treatment and a relevant comparator. \cite{Thalletal:1988, StallardTodd:2003, Schaidetal:1990} In particular, Wang \textit{et al.} develop a two-stage design that uses a non-inferiority framework for the final comparative effectiveness analysis with normally distributed outcomes.\cite{Wangetal:2017} They highlight that extending these two-stage designs to a non-inferiority framework is non-trivial due to the complex null hypothesis.\cite{Wangetal:2017} Elsewhere, Kimani \textit{et al.} developed a seamless design that incorporated dose selection based on safety and efficacy using Bayesian methods and a final comparative analysis based on frequentist hypothesis testing.\cite{Kimanietal:2009} 

The Ketodex trial, an investigator-initiated trial looking at treatments for procedural sedation within a paediatric emergency department,\cite{Poonaietal:2020} aims to determine the optimal dose combination among three alternatives and assess whether it is non-inferior to the standard of care using a binary outcome. Thus, we developed a novel trial design that uses response adaptive randomisation,\cite{Lewisetal:2013} dose response modelling for dose combinations,\cite{Caietal:2014} and a comparative effectiveness analysis from a non-inferiority perspective with binary outcomes. In contrast to the currently available methods, we used a Bayesian framework for both the dose selection and comparative effectiveness analysis. This trial also avoids a formal separation between the two study phases to maximise the time available to assess the relative efficacy of the different dose combinations.

We used Bayesian response adaptive randomisation to increase the expected allocation of patients to the optimal dose combination.\cite{Lewisetal:2013, Cellamareetal:2017} While adaptive randomisation in multi-arm trials can lead to low probabilities of selecting the true optimal treatment,\cite{WathenThall:2017} we found that the proposed adaptive randomisation scheme yielded a high probability of selecting the correct optimal treatment, based on our assumptions. Using dose response modelling for the different combination therapies then ensures that all participant data is used in the final comparative analysis.\cite{Caietal:2014} Typically, the analysis of a seamless trial requires adjustments to ensure valid construction of confidence intervals and $p$-values.\cite{Pallmannetal:2018} However, we avoided this requirement as all inference was proposed from a Bayesian perspective.\cite{Berryetal:2010}

This paper presents our novel trial design and the simulation scenarios used to determine decision thresholds for the trial adaptions and the comparative analyses.\cite{Kairallaetal:2012} All thresholds were chosen to ensure the trial had good frequentist operating characteristics and a high chance of detecting the true optimal treatment.\cite{Berryetal:2010} We also introduced a novel framework for drawing conclusions based on the study data where an inconclusive trial outcome suggests additional data should be collected. Finally, we demonstrated that this design has high Bayesian predictive power. 

\subsection{The Ketodex Trial}
Procedural sedation and analgesia (PSA) is used to facilitate the realignment of a fractured or dislocated limb without surgery in children (known as a closed reduction).\cite{Schofieldetal:2013, ChengShen:1993} Intravenous (IV) ketamine is often used to provide PSA.\cite{Schofieldetal:2013} However, IV insertion is distressing to children and their families and is often technically challenging.\cite{Cummingsetal:1996} Thus, an alternative administration method would be preferred, e.g. intranasal (IN) administration. There is limited evidence that a combination of ketamine and dexmedetomidine (ketodex) given intranasally may offer adequate sedation.\cite{Qiaoetal:2017, Bhatetal:2016} However, this combination has not been trialled in patients undergoing a closed reduction and so the Ketodex trial aims to: 
\begin{itemize}
    \item[(i)] determine a suitable combination of IN ketodex and
    \item[(ii)] compare the efficacy of IN ketodex (novel combination therapy) to IV ketamine (standard of care).
\end{itemize} As IN delivery of sedative agents is preferable to IV insertion, the Ketodex trial considers whether IN ketodex is non-inferior to IV ketamine. 

The Ketodex trial has a binary primary outcome where ``success'' is defined as a patient who is adequately sedated throughout the closed reduction procedure. Adequate sedation is defined as \begin{enumerate}
    \item[(i)] A Paediatric Sedation State Score (PSSS) of 2 or 3 for the duration of the procedure;\cite{Craveroetal:2017}
    \item[(ii)] No additional medication given for the purposes of sedation during the procedure;
    \item[(iii)] The patient does not actively resist, cry or require physical restraint to complete the procedure.
\end{enumerate} 
Patients, aged 4 - 17 (maximum 70kg), are enrolled if the physician expects to complete the procedure within five minutes. The primary outcome is measured by blinded video assessors within 24 hours of enrolment and will, thus, be used to assess the efficacy of the different ketodex combinations and the relative effectiveness of IN ketodex to IV ketamine. Clinical expertise based on a recent systematic review,\cite{Poonaietal:2020b} selected three dose combinations for IN ketodex for the trial:
\begin{enumerate}
\item A single dose of ketamine at 2 mg/kg combined with dexmedetomidine at 4 mcg/kg (2-4 ketodex)
\item  A single dose of ketamine at 3 mg/kg combined with dexmedetomidine at 3 mcg/kg (3-3 ketodex)
\item  A single dose of ketamine at 4 mg/kg combined with dexmedetomidine at 2 mcg/kg (4-2 ketodex)
\end{enumerate}
Full details of the trial conduct and outcomes are in the trial protocol.\cite{Poonaietal:2020}

\section*{Methods}
\begin{figure*}[ht]
\centering
\includegraphics[width = 0.95\textwidth]{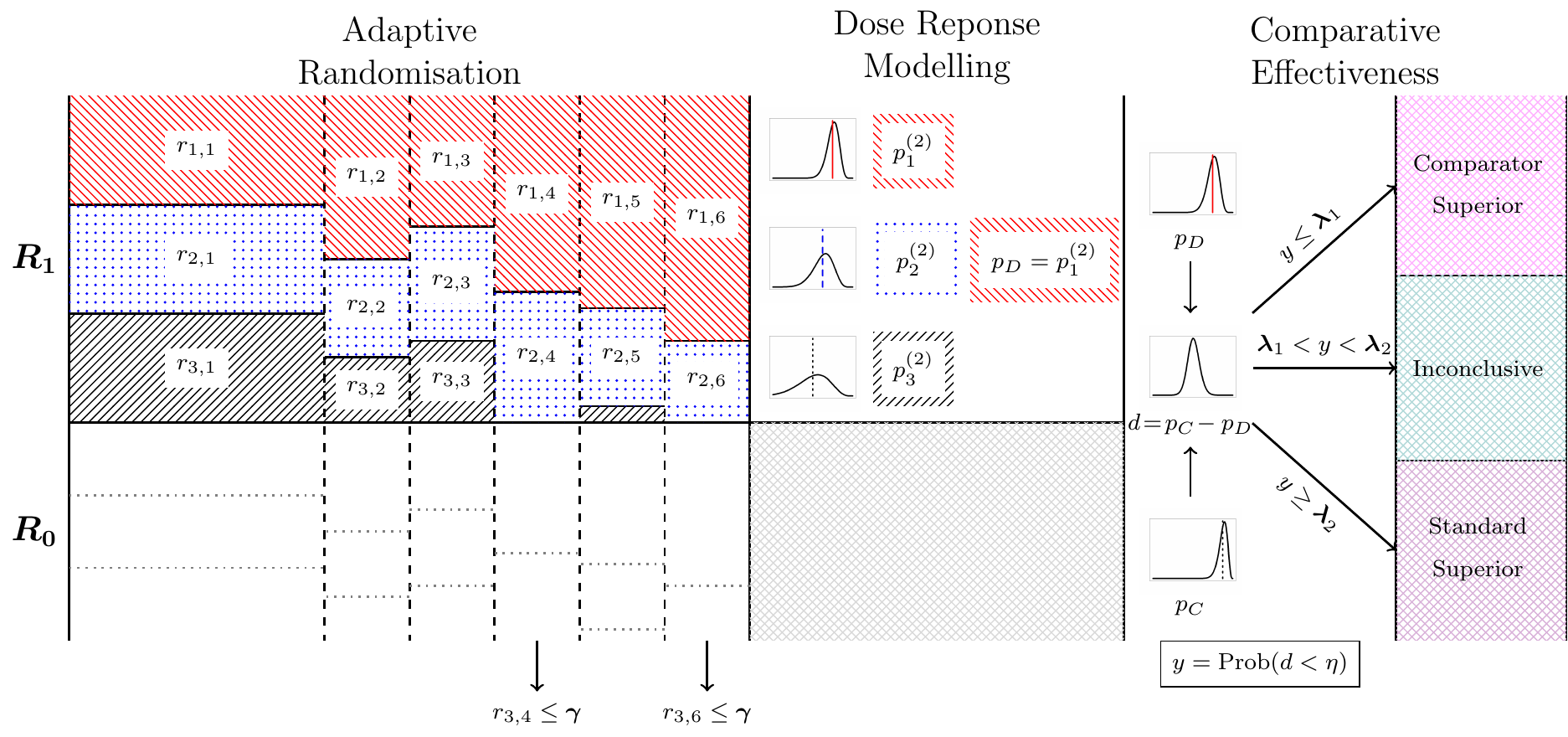}
\caption{A graphical representation of the proposed trial design. The trial design and analysis consists of three key steps. We use response adaptive randomisation for the $R_1$ proportion of patients receiving the novel combination (represented in the top half of the Figure). The $R_0$ proportion of patients who receive the standard of care are also randomised to a combination of placebo agents to maintain blinding. Within the adaptive randomisation section, $r_{i,j}$ is the proportion of patients randomised to dose combination $i$ in trial period $j$. Further, if $r_{i,j}$ is less than a threshold $\gamma$, no patients are randomised to this combination in trial period $j$. In the second step, we use Bayesian dose response modelling to determine the posterior distribution of the probability of success for each dose combination ($p_i^{(2)}$, $i = 1, 2, 3$). The optimal treatment (probability of success denoted $p_D$) is the treatment with the maximum expected probability of success, represented by the vertical lines in the top of the second panel. Finally, we compare the effectiveness of the dose combination to the effectiveness of the standard of care (probability of success denoted $p_C$) by computing the probability that the difference ($d$) in effectiveness is below a given threshold ($\eta$). Conclusions following the trial are made based on thresholds $\lambda_1$ and $\lambda_2$, chosen by simulation. Notation in bold is determined using simulation methods.}
\label{Fig:DesignPicture}
\end{figure*}

Figure \ref{Fig:DesignPicture} is a graphical representation of our novel trial design, consisting of three key steps; adaptive randomisation, dose response modelling and a comparative effectiveness analysis.

\subsection*{Overall Sample Size}
The overall sample size was determined based on pragmatic concerns and the \emph{Average Length Criterion} (ALC).\cite{JosephBelsie:1997, Caoetal:2009} ALC selects the smallest sample size for which the posterior credible interval has an average length below some fixed constant $\zeta$, to be specified. This trial design controlled the average length of the high-density posterior credible interval of the difference in effectiveness between the standard of care and the novel combination. We also used the ALC to determine the appropriate randomisation ratio between the novel combinations and the standard of care. For this overall sample size calculation, we did not consider any adaptive elements. However, our simulations ensured well-controlled error rates conditional on this sample size and the proposed adaptive design.

\subsection*{Response-Adaptive Randomisation}
We fixed the proportion of patients receiving the standard of care, to ensure enough data for the comparative effectiveness analysis, \cite{WathenThall:2017} and applied response-adaptive randomisation to increase the number of participants receiving effective dose combinations.\cite{Lewisetal:2013}  Practically, this required a two-stage randomisation procedure where participants are initially randomised to either the standard of care or the dose combinations. We define $R_0$ as the proportion of participants randomised to receive the standard of care and $R_1 = 1 - R_0$ as the proportion of participants randomised to receive the dose combinations. Participants who receive the dose combinations are then further randomised to receive a specific combination in a randomisation ratio that is updated at each interim analysis (first panel in Figure \ref{Fig:DesignPicture}). Note that, to maintain blinding in the Ketodex trial, all participants are randomised to a dose combination, even if they receive the standard of care and the dose combination would be between two placebo agents.

Formally, in the second randomisation step, the randomisation ratio across the different dose combinations $i = 1, 2, 3$ is set separately for each trial period $j = 1,\dots, J$, where $J$ is the number of phases for the adaptive randomisation (6 in Figure \ref{Fig:DesignPicture}). We denote the proportion of participants randomised to dose combination $i$, given that they are receiving the active combination, in trial period $j$, $r_{i,j}$ and, thus, the overall proportion of participants randomised to dose combination $i$ in period $j$ is $R_1 \times r_{i,j}$. 

We set $r_{i,j}$ equal to the probability that dose combination $i$ is optimal given the available evidence; \[r_{i,j} = \mbox{Prob}\left(p_i = \max_i \{p_i\}\right)\] where $p_i$ is the probability of success for the dose combination $i$. The posterior distribution of $p_i$ will be obtained conditional on the data collected in trial periods $1$ to $j$. We use the same prior distribution for $p_i$, $i=1,\dots,3$ so $r_{i, 1} = \frac{1}{3}$. To avoid randomising a small number of participants to a single arm,\cite{WathenThall:2017} we determined a value $\gamma$ such that we set $r_{i,j} = 0$ when $r_{i,j} \leq \gamma$. We then adjust the values of $r_{i,j}$ so $\sum_{i=1}^3 r_{i,j} = 1$. Note that all trial data will be included in the final analysis and a dose combination can be reinstated in the randomisation scheme even if it was excluded previously (Figure \ref{Fig:DesignPicture}, trial periods 4 and 5).

The practicalities of the Ketodex trial meant that the second step randomisation is undertaken without knowledge of the treatment group assignment from the first randomisation. This means that we cannot fix the exact proportion of participants receiving dose combination $i$ in period $j$ to $R_1 \times r_{i,j}$. Therefore, assuming that the number of participants to be enrolled in trial period $j$ is $N_j$, the number of participants randomised to each active dose combination $i$ will be a random variable \[M_{i, j} \sim \mbox{Binomial}(N_j r_{i,j}, R_1).\] 

\subsection*{Dose Response Modelling}
The primary effectiveness analysis will use a dose response model to estimate the probability of success for each of the three dose combinations. \cite{Kimanietal:2009,Caietal:2014} To achieve this, note that participants who are not adequately sedated can either be over-sedated (PSSS score of 0 or 1) or under-sedated (PSSS score of 4 or 5).\cite{Craveroetal:2017} We assume that the probability of over- and under-sedation can each be modelled with a monotonic log-logistic dose response model based on expert opinion. We then use a multinomial distribution to make inferences about the probability of adequate sedation. 

Let $\bm X_i$, $i = 1, 2, 3$, be a 3-vector containing the number of patients who experience under, adequate and over-sedation from the $N_i = \sum_{j = 1}^J M_{i, j}$ participants who receive dose $i$; \[ X_i \sim \mbox{Multinomial}(N_i, \bm p_i),\] with  $\bm p_i = c(p_i^{(1)}, p_i^{(2)}, p_i^{(3)})'$, \[\mbox{logit}(p^{(1)}_i) = \beta_0 +\beta_1 \log(A_i) +\beta_2 \log(B_i),\] \[\mbox{logit}(p^{(3)}_i) = \beta_a +\beta_b \log(A_i) +\beta_c \log(B_i),\] and $A_i$ is dose for drug A (ketamine) and $B_i$ is the dose for drug B (dexmedetomidine). We have not included interaction terms as they cannot be reliably estimated and models without interactions perform well in dose finding studies.\cite{WangIvanova:2005}

Using these dose response models, we can determine the posterior for $p_i^{(2)}$, $i = 1, 2, 3$, the probability of adequate sedation for each dose combination. The optimal dose combination is the combination with the maximum expected value of $p_i^{(2)}$. In Figure \ref{Fig:DesignPicture}, the expected values of $p_i^{(2)}$ are shown using vertical lines, with the highest expected value associated with the first dose combination. For the comparative effectiveness analysis, we denote the probability of a success for the optimal dose combination as $p_D$, i.e., in Figure \ref{Fig:DesignPicture}, $p_D = p_1^{(2)}$.

\subsection*{Comparative Effectiveness Analysis}

The comparative effectiveness analysis compares the probability of success for standard of care, $p_C$, to $p_D$. In a Bayesian framework, we compute the posterior distribution of $d = p_C - p_D$ and then calculate the probability that $d$ is greater than a pre-specified value $\eta$; $y = P(d \geq \eta)$. A superiority trial would set $\eta = 0$ but, the Ketodex trial aims to determine whether IN ketodex is non-inferior to IV ketamine and, therefore, $\eta$ is the non-inferiority margin of $0.178$. This non-inferiority margin is the average from the responses of 204 clinicians in surveys undertaken by the Ketodex team. Small values of $y$ are evidence of non-inferiority.

The proposed trial has three potential outcomes (Figure \ref{Fig:DesignPicture}):
\begin{enumerate}
    \item The optimal dose combination is superior/non-inferior, 
    \item The trial is inconclusive or,
    \item The standard of care is superior.
\end{enumerate}
The trial conclusion is made using two thresholds $\lambda_1$ and $\lambda_2$, chosen by simulation to control the trial errors rates, as required for Bayesian designs.\cite{Berryetal:2010} Specifically, if $y \leq \lambda_1$ then we conclude that the optimal dose combination is superior/non-inferior to the standard of care. If $y \geq \lambda_2$, then we conclude that the standard of care is superior to the optimal dose combination. As $\lambda_1 + \lambda_2 < 1$, any values of $y$ between $\lambda_1$ and $\lambda_2$ will be deemed inconclusive, i.e., the current data are insufficient to determine whether the novel combination or the standard of care is superior. This conclusion encourages the collection of more data to deliver definitive conclusions.

\subsection*{Simulation Scenarios}
We used simulations to develop and evaluate our trial design. To achieve this, we used four different simulation settings to:
\begin{enumerate}
    \item Evaluate the ALC to determine the overall sample size and value of $R_0$; the proportion of participants randomised to the standard of care.
    \item Determine the value of $\gamma$; the threshold under which we drop a dose combination from the randomisation procedure.
    \item Determine the values of $\lambda_1$ and $\lambda_2$; the thresholds for concluding non-inferiority of the optimal dose combination or the standard of care.
    \item Compute the predictive power of the trial.
\end{enumerate} 
These simulation scenarios were undertaken sequentially, i.e., the overall sample size was determined and then used as the sample size throughout the remaining simulation scenarios. The following sections outline the parameters of these four simulation settings and the criteria used draw conclusions from each simulation for the Ketodex trial.

\subsubsection*{Determining the Overall Sample Size}
We set $\zeta = 0.07$ and controlled the length of the 95\% highest density posterior credible interval. The value for $\zeta$ represents a posterior credible interval that is six times shorter than the prior credible interval and was chosen considering the budget and time constraints that limited our maximum recruitment.\cite{Caoetal:2009}

ALC determined the overall trial sample size and $R_0$, the proportion of participants randomised to the standard of care. We considered four values for $R_0$, 0.2, 0.3, 0.4 and 0.5, and computed the ALC for each $R_0$ for sample sizes increasing in increments of 10 from 350 to 500. We selected the smallest sample size that respects the ALC and then, for that sample size, we selected the value of $R_0$ that leads to the most balanced trial, provided the ALC is respected.

For each sample size and value of $R_0$, we simulated 2000 datasets from the prior-predictive distribution of the data, using the priors defined below. For each dataset, we obtained 2000 simulations from the posterior distribution of $p_C$ and $p_D$ and computed the highest density posterior 95\% credible interval.\cite{MeredithKruschke:2018} We estimated their average length for each sample size and value of $R_0$ across all 2000 prior-predictive datasets. 

\subsubsection*{Determining $\gamma$ for the adaptive randomisation}
We used simulations to determine the value of $\gamma$, the threshold for dropping a given dose combination from randomisation in a specific trial period. We considered values between 0.05 and 0.3, increasing in increments of 0.05, with 0.3 chosen as the maximum because an even randomisation ratio would have 0.33 randomised to each arm. We fixed the probability of adequate sedation for the three dose combinations at $0.93, 0.88$ and $0.83$, as $0.93$ is the success rate seen in a previous trial of IN Ketodex \cite{Bhatetal:2016}. We then calculated the number of participants randomised to each treatment option for each value of $\gamma$. Across 7000 simulations, we estimated the probability of randomising the highest number of participants to the dose with probability of adequate sedation equal to $0.93$. We selected the value of $\gamma$ that maximises this probability. If two values for $\gamma$ gave the same probability, we chose the smallest threshold $\gamma$. This maximises the amount of information collected for the optimal treatment, if the incorrect optimal treatment were selected.

We used 7000 simulations as it gives a greater than 99\% chance of estimating the probability of 0.8 to 2 decimal places, the accuracy chosen for all analyses in this manuscript. 

\subsubsection*{Thresholds for Comparative Effectiveness Analysis}
Based on the adaptive randomisation scheme finalised in the previous section, we used simulation to determine the decision thresholds for the comparative effectiveness analysis, $\lambda_1$ and $\lambda_2$. As $\lambda_1$ controls the type I error of the trial, we set the probability of adequate sedation for IN Ketodex $p_C = 0.97$,\cite{Kannikeswaranetal:2016} and the probability of adequate sedation for the optimal dose combination $p_D = p_C - \eta = 0.792$. We then selected $\lambda_1$ such that 5\% of the trials incorrectly conclude non-inferiority. For $\lambda_2$, we set $p_D = 0.78$, and undertook the same trial simulation process, specifying that 50\% of the simulated trials should declare superiority for the standard of care. We set 3-3 ketodex as the optimal treatment, based on expert guidance, and the probability of adequate sedation of 4-2 ketodex to $p_D - 0.05$ and 2-4 ketodex as $p_D - 0.1$. Expert guidance specified that high doses of dexmedetomidine could lead to over sedation. Thus, we assumed that the proportion of people that were over-sedated, among those who were inadequately sedated, was 0.01, 0.1 and 0.2 for ketodex 4-2, 3-3 and 2-4, respectively. Finally, we simulated 7000 trials with the proposed comparative effectiveness analysis using 7000 posterior simulations for $\bm p_i$, $i = 1, 2, 3$, and $p_C$ for each trial.

Finally, to understand the design characteristics further, we evaluated the probability of each trial outcome, for 8 different values of $p_D$; $p_D = 0.93, 0.9, 0.87, 0.85, 0.83, 0.792, 0.78$ and $0.75$, using the specified values of $\lambda_1$ and $\lambda_2$. Table \ref{tab:Scenarios} outlines the values set for each probability of interest in all these scenarios.

\begin{table*}[ht]
    \centering
    \begin{tabular}{p{1.4cm}|c|c|c|c|c|c|c}
    \multirow{2}{*}{\parbox{1.4cm}{Scenario Number}} & \multicolumn{3}{|p{3.6cm}|}{Probability of Adequate Sedation} & \multicolumn{3}{|p{3.6cm}|}{Probability of Over-Sedation} & \multirow{2}{*}{\parbox{4.2cm}{Number of Non-Inferior Dose Combinations}} \\ \cline{2-7}
    & 3-3 & 4-2 & 2-4 & 3-3 & 4-2 & 2-4 & \\ \hline
        1 & 0.93 & 0.88 & 0.83 & 0.007 & 0.0012 & 0.034 & 3  \\
        2 & 0.90 & 0.85 & 0.80 & 0.010 & 0.0015 & 0.040 & 3  \\
        3 & 0.87 & 0.83 & 0.77 & 0.013 & 0.0017 & 0.046 & 2  \\
        4 & 0.85 & 0.80 & 0.75 & 0.015 & 0.0020 & 0.050 & 2  \\
        5 & 0.83 & 0.78 & 0.73 & 0.017 & 0.0022 & 0.054 & 1  \\
        6 & 0.792 & 0.742 & 0.692 & 0.021 & 0.0026 & 0.062 & 0  \\
        7 & 0.78 & 0.73 & 0.68 & 0.022 & 0.0027 & 0.064 & 0  \\
        8 & 0.75 & 0.70 & 0.65 & 0.025 & 0.0030 & 0.070 & 0  \\
    \end{tabular}
    \caption{The values set for the probability of adequate and over-sedation in the eight scenarios considered to assess the operating characteristics of the Ketodex trial. The probabilities are given for each of the three dose combinations 3-3 ketodex, 4-2 ketodex and 2-4 ketodex. In each scenario, we highlight the number of dose combinations that are non-inferior to IV ketamine.}
    \label{tab:Scenarios}
\end{table*}

\subsubsection*{Bayesian Predictive Power}
We calculated the expected probability of a conclusive trial, i.e., concluding superiority of the novel combination or the standard of care, using Bayesian predictive power. We took 2000 simulations from the prior predictive distributions of $p_C$ and $p_D$, using the priors outlined below. We considered three scenarios where we varied the relative risk of adequate sedation for the three dose combinations relative to our prior beliefs about $p_D$ (outlined in Table \ref{tab:ScenariosBayes}). Scenario A set the relative risk to 0.9, 0.95 and 1 for 4-2, 2-4 and 3-3 ketodex, respectively. Similarly, Scenario B used 0.95, 0.98 and 1 and Scenario C used 0.95, 1 and 1.05. We assumed that the proportion of inadequately sedated patients who were over-sedated was 0.01, 0.1 and 0.2 for 4-2, 3-3 and 2-4 ketodex, respectively. We used 2000 posterior simulations for $\bm p_i$, $i = 1, 2, 3$, and $p_C$ to perform the comparative effectiveness analysis for each prior predictive sample.

\begin{table}[ht]
    \centering
    \begin{tabular}{p{1.4cm}|c|c|c|c|c|c}
    \multirow{2}{*}{\parbox{1.4cm}{Scenario Letter}} & \multicolumn{3}{p{3.2cm}|}{Relative Risk of Adequate Sedation} & \multicolumn{3}{|p{6.1cm}}{Proportion of over-sedated participants among the inadequately sedated participants} \\ \cline{2-7}
    & 3-3 & 4-2 & 2-4 & 3-3 & 4-2 & 2-4 \\ \hline
        A & 1 & 0.9 & 0.95 & 0.1 & 0.01 & 0.2  \\
        B & 1 & 0.95 & 0.98 & 0.1 & 0.01 & 0.2  \\
        C & 1.05 & 0.95 & 1 & 0.1 & 0.01 & 0.2  \\
    \end{tabular}
    \caption{The scenarios considered to estimate the Bayesian predictive power of the Ketodex trial. We specified the relative risk of adequate sedation compared to our prior beliefs about the probability of adequate sedation. Secondly, we specified the proportion of over-sedated participants, among those who were inadequately sedated, for each of the three dose combinations. In this case, using a prior distribution for the probability of adequate sedation means that we cannot display the probability of over-sedation. Both key quantities are given for each of the three dose combinations 3-3 ketodex, 4-2 ketodex and 2-4 ketodex.}
    \label{tab:ScenariosBayes}
\end{table}

\subsection*{Prior Specification}
We specified priors for $p_C$, the probability of success for the standard of care, $\beta_0$, $\beta_1$, $\beta_2$,  $\beta_a$, $\beta_b$, and $\beta_c$, the parameters of the log-logistic dose-response models for under- and over-sedation. For the prior predictive analyses, we also specified a prior for $p_D$, the probability of success for the optimal dose combination, directly. We used either published evidence or minimally informative priors.

For IV ketamine, Kannikeswaran \textit{et al.}~had a 97\% success rate with ketamine dosed at 1.5 mg/kg, as proposed in the Ketodex trial.\cite{Kannikeswaranetal:2016} To account for differences between this trial and the Ketodex trial, we down-weighted this information to an effective sample size of 16; \[p_C \sim \mbox{Beta}(15.6,0.44).\] Bhat \textit{et al.}~published a trial in which 2 out of 27 participants were inadequately sedated with IN Ketodex dosed at 1 mg/kg of ketamine and 2 $\mu$g/kg of dexmedetomidine.\cite{Bhatetal:2016} To account for substantial differences between the two trials, including in dosing and the setting, we discount this information to an effective sample size of 6.5; \[ p_D \sim \mbox{Beta}(6.25, 0.25).\] The effective sample size of the prior for $p_D$ means that \emph{a priori} there is a 90\% chance that IN ketodex is non-inferior to IV ketamine, using the non-inferiority margin $\eta = 0.178$.

For the regression coefficients, we used non-central Student t-distributions with precision 0.001 and degrees of freedom 1.\cite{Gelmanetal:2008} We set the mean for $\beta_1$, $\beta_2$, $\beta_b$ and $\beta_c$ to be 0, as we have minimal information on the dose response. For $\beta_0$ and $\beta_a$, we set their means such that 5\% of participants are expected to be under-sedated and 2\% of participants are expected to be over-sedated. Thus, the prior means of $p_i^{(2)}$, $i=1, 2, 3$ are $0.93$, the success rate observed in the literature.\cite{Bhatetal:2016}  

\section{Results}

\subsection*{Average Length Criterion}
\begin{figure}[ht]
    \centering
    \includegraphics[width = \columnwidth]{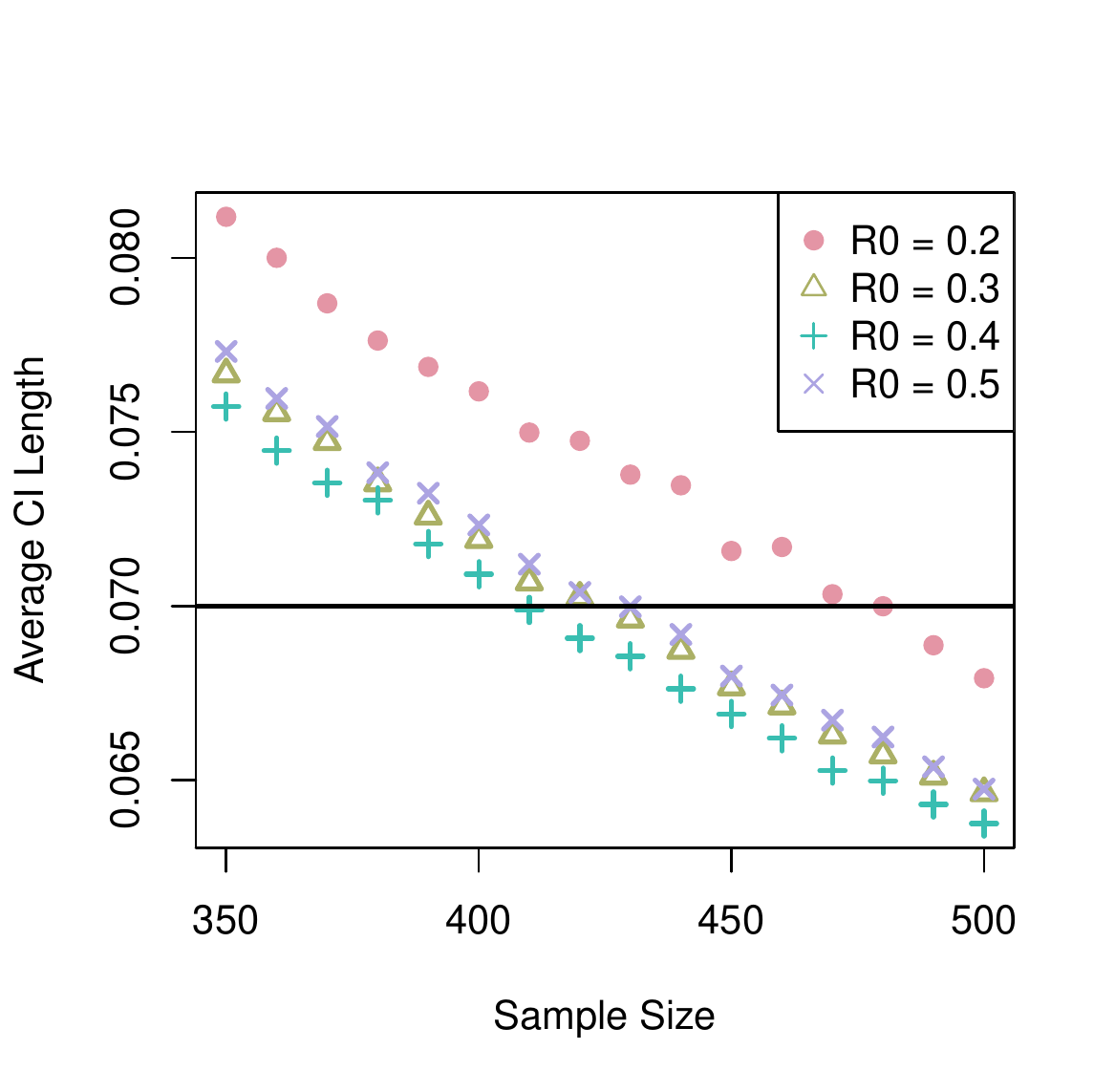}
    \caption{The average length of the posterior highest density credible interval for the four alternative values of $R_0$, the proportion of participants randomised the comparator, across sample size. A black line represents the threshold of 0.07.}
    \label{fig:ALC}
\end{figure}

Figure \ref{fig:ALC} displays the results from the ALC analysis. The ALC is respected with a sample size of 410 and $R_0 = 0.4$. For all sample sizes, $R_0 = 0.4$ results in the shortest credible intervals. With this sample size, we structured the interim analyses using pragmatic concerns. The first interim analysis will take place after an expected enrolment of 30 participants for each dose combination to ensure sufficient data is collected before changing the randomisation ratio.\cite{Thorlundetal:2018} Thus, the first interim analysis will take place once 150 participants have been enrolled. Further updates of the randomisation ratio will take place at intervals of 50 participants, i.e., at 150, 200, 250, 300, 350, before the final comparative effectiveness analysis at 410.

\subsection*{Adaptive Randomisation: Choosing $\gamma$}
Table \ref{tab:Gamma} displays the probability of randomising the highest number of people to the true optimal dose combination. This probability is maximised when $\gamma = 0.05, 0.2, 0.25, 0.3$, with an 83\% chance of randomising the most people to the optimal treatment. Thus, we chose the smallest value; $\gamma = 0.05$ and expect to enrol 137 patients to the optimal treatment with an 95\% interval between 56 and 190.

\begin{table}[ht]
    \centering
    \begin{tabular}{l|l|l|l|l|l|l}
        $\gamma$ &  0.05 & 0.1 & 0.15 & 0.2 & 0.25 & 0.3\\ \hline
        Probability & 0.83 & 0.82 & 0.82 & 0.83 & 0.83 & 0.83 \\
    \end{tabular}
    \caption{The probability of randomising the greatest number of patients to the best dose combination, conditional on different thresholds for dropping arms with limited evidence of effectiveness ($\gamma$).}
    \label{tab:Gamma}
\end{table}

\subsection*{Comparative Effectiveness Analysis: Choosing $\lambda_1$ and $\lambda_2$}

Based on the overall sample size and the values for $R_0$ and $\gamma$, $\lambda_1 = 0.037$ ensures a type I error of 5\% size and $\lambda_2 = 0.608$ ensures that 50\% of trials declare superiority for IV ketamine when $p_D = 0.78$. Based on these thresholds, Table \ref{tab:Power} gives the probability of each trial outcome for all the considered scenarios. The probability of concluding non-inferiority is very high at the prior mean of 0.93 and remains above 90\%, provided the true probability of success for the optimal dose combination is over 0.9. The probability of an inconclusive trial is high for $p_D$ close to the non-inferiority threshold. The trial has a higher probability of being inconclusive if $p_D$ is above the non-inferiority boundary of 0.792. For example, the probability of an inconclusive trial is approximately 0.5 for both $p_D = 0.85$ and $p_D = 0.78$ (table \ref{tab:Power}) but 0.85 is further from the non-inferiority boundary than 0.78. 

The probability of both declaring non-inferiority and selecting the correct optimal treatment is lower than the baseline probability of declaring non-inferiority. However, it is above 77\% when $p_D$ is above 0.9. This probability is small as we approach the non-inferiority margin but not substantially reduced from the underlying probability of declaring non-inferiority.

\begin{table*}[ht]
    \centering
    \begin{tabular}{|c|c|c|c|c|c|}
    \hline
   & & \multicolumn{4}{c|}{Probability of}  \\ \cline{3-6}
    \parbox{1.6cm}{Scenario Number} & $p_D$ & Non-inferiority & Inconclusive & Superiority & \parbox{2cm}{Non-Inferiority and Correct Optimal Treatment}\\ \hline
    1& 0.93 & $> 0.99 $ &  $< 0.01$ & $< 0.01$ & 0.84 \\
    2&0.90 & $0.93$ & 0.07 & $< 0.01$ & 0.77 \\
    3&0.87 & 0.71 &  0.29 & $< 0.01$ & 0.61\\
    4&0.85 & 0.48 & 0.50 & 0.02 & 0.42\\
    5&0.83 & 0.27 & 0.66 & 0.07 & 0.25\\
    6&0.792 & 0.05 &  0.58 & 0.37 & - \\
    7&0.78 & 0.02 &  0.48 & 0.50 & -\\
    8&0.75 & $< 0.01$ &  0.22 & 0.78 & -\\ \hline
    \end{tabular}
    \caption{The operating characteristics of the novel Bayesian design for the Ketodex trial by simulation number (Table \ref{tab:Scenarios}). We report the probability of each trial outcome, Non-Inferiority, Inconclusive and Superiority for each of the 8 scenarios we consider. We also report the probability that the trial simultaneously selects the correct optimal treatment and concludes non-inferiority. We do not report this value when all three dose combinations are inferior (scenarios 6, 7 and 8). The probability of effectiveness for the optimal dose combination is listed in the $p_D$ column. The complete specification of all probabilities for each scenario is outlined in Table \ref{tab:Scenarios}.}
    \label{tab:Power}
\end{table*}

\subsection*{Predictive Power}
The Bayesian predictive power of the Ketodex trial, i.e., the prior probability that the trial is conclusive, is 0.92, 0.92 and 0.94 for scenarios A, B and C, respectively (see Table \ref{tab:ScenariosBayes}). The predictive power is higher for scenario C as the optimal combination was assumed more effective than the combination seen in the literature. In all scenarios, the predictive power is over 90\%, which is higher than the prior probability that IN ketodex is superior to IV ketamine. This is because we can conclude that IV ketamine is superior to IN ketodex. The predictive power of declaring non-inferiority is 0.83, 0.84 and 0.88 for scenarios A, B and C, respectively.

\section*{Discussion}
We developed a novel Bayesian trial design that evaluates the comparative effectiveness of a novel combination therapy in a non-inferiority framework and determines the optimal dose combination. We used response adaptive randomisation to increase the number of participants receiving the higher performing dose combinations and dose-response modelling to increase the power of the comparative effectiveness analysis. This trial minimises the administrative burden of evaluating novel combination therapies and, although it is applied to a non-inferiority setting, can easily be adapted to evaluate superiority of the novel combination. 

Our design has a high chance of reaching the dual study aims in settings where the probability of effectiveness for the dose combination is consistent with previous studies. Note that the Ketodex trial is a non-inferiority trial with a large non-inferiority margin and, thus, the sample size requirements for this design may increase substantially as the non-inferiority margin gets smaller. Furthermore, the components of this novel design must be re-estimated in alternative settings, which may change the operating characteristics. However, we have included all code for this design in the supplementary material to facilitate the reuse of this design in other settings. 

A limitation of this trial was the decision to restrict ourselves to three dose combinations, rather than investigate all possible dose combination pairs. This restriction was made for pragmatic reasons based on clinical judgement, informed by the literature.\cite{Poonaietal:2020b} However, there is a possibility that the optimal dose combination is not included in the combinations investigated in this trial.

This novel design allowed for an inconclusive trial based on posterior probabilities. To compute these probabilities, we must have a one-sided test rather than a point hypothesis. Thus, an adaptation of this decision rule would be required for two-sided tests. Nonetheless, we included the possibility of an inconclusive trial outcome based on the Bayesian trial analysis to encourage the collection of further information, past the initial completion date of the trial, if additional data were required to assess the comparative effectiveness of the two treatments.

Finally, this design and decision making framework could be used for a seamless dose finding phase II/III trial for a novel drug. However, this requires a formal assessment of safety for the novel intervention.\cite{Kimanietal:2009} The advantage of dose combination studies is that safety is well understood so a statistical assessment of safety may not be required, as in the Ketodex trial, although safety should always be considered by the data safety monitoring board. 

\subsection*{Conclusion}
We developed a novel trial design to undertake dose finding and comparative effectiveness analysis that had good statistical properties and respected the time and resource constraints of an investigator initiated trial.

\subsection*{Acknowledgements}
The authors would like to acknowledge the KidsCAN PERC Innovative Pediatric Clinical Trials Ketodex Study Team, the iPCT SPOR administrative staff and our patient partners, who provided valuable support and input on the full Ketodex study design and documents. We would also like to thank an associate editor for their insightful comments that have greatly improved this manuscript.

\subsection*{Funding}
\textbf{The author(s) disclosed receipt of the following financial support for the research, authorship, and/or publication of this article}: This work is supported by an Innovative Clinical Trials Multi-year Grant from the Canadian Institutes of Health Research (funding reference number MYG-151207; 2017 – 2020), as part of the Strategy for Patient-Oriented Research and the Children’s Hospital Research Institute of Manitoba (Winnipeg, Manitoba), the Centre Hospitalier Universitaire Sainte-Justine (Montreal, Quebec), the Department of Pediatrics, University of Western Ontario (London, Ontario), the Alberta Children’s Hospital Research Institute (Calgary, Alberta), the Women and Children’s Health Research Institute (Edmonton, Alberta), the Children’s Hospital of Eastern Ontario Research Institute Inc. (Ottawa, Ontario), and the Hospital for Sick Children Research Institute (Toronto, Ontario). This study is sponsored by The Governors of the University of Alberta (Suite 400, 8215 – 112 Street, Edmonton, Alberta, Canada T6G 2C8). Neither the study sponsor nor funders have any role in the collection, management, analysis, or interpretation of data; writing of the report; or the decision to submit the report for publication. Additional support was received from the Physicians Services Incorporated Foundation, Academic Medical Organization of Southwestern Ontario, Ontario Ministry of Economic Development, Job Creation and Trade, and the Children’s Health Foundation of the Children’s Hospital, London Health Sciences Foundation.

\subsection*{Declaration of Conflicting Interests} None declared.

\bibliographystyle{SageV}
\bibliography{bib}

\end{document}